\documentclass[twoside,twocolumn,9pt]{article}
\usepackage{extsizes}
\usepackage[super,sort&compress,comma]{natbib} 
\usepackage[version=3]{mhchem}
\usepackage[left=1.5cm, right=1.5cm, top=1.785cm, bottom=2.0cm]{geometry}
\usepackage{balance}
\usepackage{mathptmx}
\usepackage{sectsty}
\usepackage{graphicx} 
\usepackage{lastpage}
\usepackage[format=plain,justification=justified,singlelinecheck=false,font={stretch=1.125,small,sf},labelfont=bf,labelsep=space]{caption}
\usepackage{float}
\usepackage{fancyhdr}
\usepackage{fnpos}
\usepackage[english]{babel}
\addto{\captionsenglish}{%
  
}
\usepackage{array}
\usepackage{droidsans}
\usepackage{charter}
\usepackage[T1]{fontenc}
\usepackage[usenames,dvipsnames]{xcolor}
\usepackage{setspace}
\usepackage[compact]{titlesec}
\usepackage{hyperref}

\usepackage{epstopdf}

\definecolor{cream}{RGB}{222,217,201}

\begin{document}

\pagestyle{fancy}
\thispagestyle{plain}
\fancypagestyle{plain}{
\renewcommand{\headrulewidth}{0pt}
}

\makeFNbottom
\makeatletter
\renewcommand\LARGE{\@setfontsize\LARGE{15pt}{17}}
\renewcommand\Large{\@setfontsize\Large{12pt}{14}}
\renewcommand\large{\@setfontsize\large{10pt}{12}}
\renewcommand\footnotesize{\@setfontsize\footnotesize{7pt}{10}}
\makeatother

\renewcommand{\thefootnote}{\fnsymbol{footnote}}
\renewcommand\footnoterule{\vspace*{1pt}%
\color{cream}\hrule width 3.5in height 0.4pt \color{black}\vspace*{5pt}} 
\setcounter{secnumdepth}{5}

\makeatletter 
\renewcommand\@biblabel[1]{#1}            
\renewcommand\@makefntext[1]%
{\noindent\makebox[0pt][r]{\@thefnmark\,}#1}
\makeatother 
\renewcommand{\figurename}{\small{Fig.}~}
\sectionfont{\sffamily\Large}
\subsectionfont{\normalsize}
\subsubsectionfont{\bf}
\setstretch{1.125} 
\setlength{\skip\footins}{0.8cm}
\setlength{\footnotesep}{0.25cm}
\setlength{\jot}{10pt}
\titlespacing*{\section}{0pt}{4pt}{4pt}
\titlespacing*{\subsection}{0pt}{15pt}{1pt}

\fancyfoot{}
\fancyfoot[LO,RE]{\vspace{-7.1pt}\includegraphics[height=9pt]{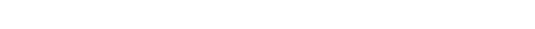}}
\fancyfoot[CO]{\vspace{-7.1pt}\hspace{13.2cm}\includegraphics{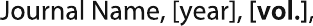}}
\fancyfoot[CE]{\vspace{-7.2pt}\hspace{-14.2cm}\includegraphics{head_foot/RF}}
\fancyfoot[RO]{\footnotesize{\sffamily{1--\pageref{LastPage} ~\textbar  \hspace{2pt}\thepage}}}
\fancyfoot[LE]{\footnotesize{\sffamily{\thepage~\textbar\hspace{3.45cm} 1--\pageref{LastPage}}}}
\fancyhead{}
\renewcommand{\headrulewidth}{0pt} 
\renewcommand{\footrulewidth}{0pt}
\setlength{\arrayrulewidth}{1pt}
\setlength{\columnsep}{6.5mm}
\setlength\bibsep{1pt}

\makeatletter 
\newlength{\figrulesep} 
\setlength{\figrulesep}{0.5\textfloatsep} 

\newcommand{\topfigrule}{\vspace*{-1pt}%
\noindent{\color{cream}\rule[-\figrulesep]{\columnwidth}{1.5pt}} }

\newcommand{\botfigrule}{\vspace*{-2pt}%
\noindent{\color{cream}\rule[\figrulesep]{\columnwidth}{1.5pt}} }

\newcommand{\dblfigrule}{\vspace*{-1pt}%
\noindent{\color{cream}\rule[-\figrulesep]{\textwidth}{1.5pt}} }

\makeatother

\twocolumn[
  \begin{@twocolumnfalse}
{\includegraphics[height=30pt]{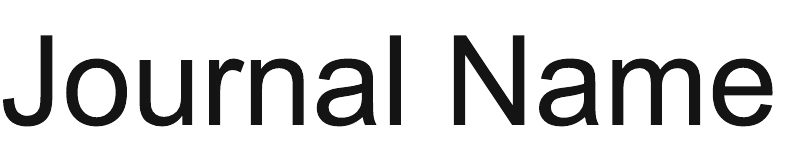}\hfill\raisebox{0pt}[0pt][0pt]{\includegraphics[height=55pt]{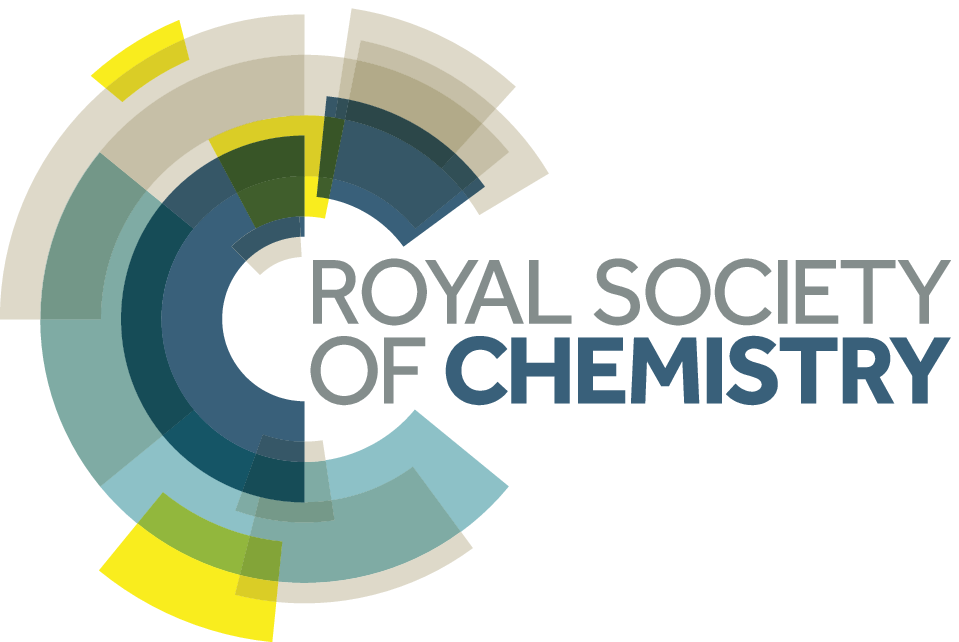}}\\[1ex]
\includegraphics[width=18.5cm]{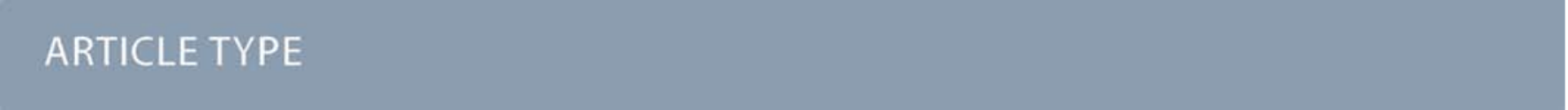}}\par
\vspace{1em}
\sffamily
\begin{tabular}{m{4.5cm} p{13.5cm} }

\includegraphics{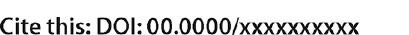} & \noindent\LARGE{\textbf{Correlations between helicity and optical losses within general electromagnetic scattering theory}} \\
\vspace{0.3cm} & \vspace{0.3cm} \\

 & \noindent\large{Jon Lasa-Alonso,\textit{$^{a, b}$} Jorge Olmos-Trigo,\textit{$^{b}$} Aitzol Garc\'ia-Etxarri,\textit{$^{b, c}$} and Gabriel Molina-Terriza\textit{$^{a, b, c}$}} \\

\includegraphics{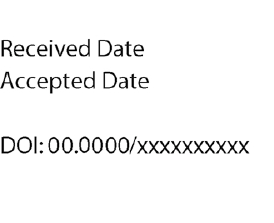} & \noindent\normalsize{Helicity preserving nanostructures and metasurfaces have been recently proposed as suitable candidates to enhance spectroscopic features of chiral molecules. With this in mind, we highlight that losses in the constituent nonmagnetic materials dramatically affect the possibility of constructing structures which conserve helicity. We first present a general procedure that permits the evaluation of the normalized helicity expectation value, $\langle\hat{\Lambda}\rangle$, i.e. the observable that permits the identification of helicity preserving scatterers. We then apply this procedure to the case of a chiral sphere, which in an orientation averaged picture can capture the optical response of chiral inorganic nanostructures, obtaining a widely applicable analytical expression of $\langle\hat{\Lambda}\rangle$ for this type of objects. Finally, we numerically show that optical losses impose an upper bound to the helicity expectation value on nonmagnetic core-shells and chiral spheres.} \\

\end{tabular}

 \end{@twocolumnfalse} \vspace{0.6cm}

  ]

\renewcommand*\rmdefault{bch}\normalfont\upshape
\rmfamily
\section*{}
\vspace{-1cm}


\footnotetext{\textit{$^{a}$~Centro de F\'isica de Materiales, Paseo Manuel de Lardizabal 5, 20018 Donostia, Spain.}}
\footnotetext{\textit{$^{b}$~Donostia International Physics Center, Paseo Manuel de Lardizabal 4, 20018 Donostia, Spain. }}
\footnotetext{\textit{$^{c}$~IKERBASQUE, Basque Foundation for Science, Mar\'ia D\'iaz de Haro 3, 48013 Bilbao, Spain.}}



The study of chirality in its various forms is a cornerstone of modern sciences. In particular, most of the biomolecules which are essential for the biological processes are chiral, i.e. they do come in two different enantiomers, which are atomically identical, but can only match to each other after applying rotations and at least a mirror reflection transformation. One way of distinguishing these two enantiomers is by their ability to absorb or delay differently the left and right circular polarizations of light. Thus, the techniques related to optical chiral sensing and circular dichroism spectroscopy, are extremely important to several branches of the chemical and biological sciences \cite{Barron, Rodger, ReviewChiral, QuantumLeap}. Unfortunately, the molecular optical rotatory power and the circular dichroism are typically small. In order to increase the light matter interaction, one can resort to highly focused beams or to the use of techniques in near field optics. In both cases, the description of the polarization of the electromagnetic field gets rather convoluted and light cannot be described anymore as a simple circularly polarized field. The helicity of light provides a solution to this problem. Optical helicity can be considered as the the nonparaxial generalization of the notion of planewave polarization~\cite{Calkin}. It is a magnitude that has gained attention in the last few years within the field of nanophotonics \cite{Tischler2014, LisaSymmetry}. In particular, this quantity is intimately related with the electromagnetic duality symmetry, i.e. that free-space Maxwell's equations are invariant under the exchange of the electric and magnetic fields \cite{PRLMolina}. Also, this symmetry can be restored in macroscopic Maxwell's equations for materials fulfilling $\varepsilon = \mu$, where $\varepsilon$ is the electric permittivity, and $\mu$ is the magnetic permeability of the sample. The symmetry between the electric and magnetic fields is also at the core of the theoretical descriptions of circular dichroism \cite{CondonReview}. On a different basis, Cohen and Tang, between others \cite{tang_cohen, tang_cohen_science, PRB}, gave an alternative theoretical measure of the circular dichroism of chiral matter under very general illumination conditions through the local density of electromagnetic chirality. Soon after, this quantity proved to be intimately connected with the helicity of light \cite{LisaSymmetry}. In fact, it has been proposed that dual particles are the adequate building blocks to enhance circular dichroism.

Light scattering on dual particles has some interesting properties. For example, on cylindrically symmetric  scatterers, the restoration of the duality symmetry implies the absence of backscattered light~\cite{ZambranaKerker}. This, provides a fundamental theoretical description of the so-called first Kerker condition~\cite{Kerker}. Helicity conservation, in principle, requires the existence of magnetic materials ($\mu \neq 1$). However, natural magnetic materials at optical frequencies do not exist, and hence, both helicity conservation and the absence of backscattered light in principle should not be observable in this spectral regime. Nevertheless, this situation changed with the advent of high refractive index (HRI) nanostructured materials that present an electric and magnetic dipolar response in the visible~\cite{strong_magnetic}. When both dipolar modes oscillate in-phase and with equal amplitudes, the first Kerker condition is satisfied~\cite{nieto2011angle} and, hence, helicity preserving nanoparticles can be built from nonmagnetic constituent materials with $\mu=1$. As a result, the zero optical backscattering condition could be experimentally measured in the visible spectral regime for HRI Si nanospheres~\cite{fu2013directional}. Since then, the conservation of helicity in nonmagnetic particles has been widely employed by different branches of optics, ranging from enhanced errors in optical localization~\cite{OpticalMirages} or isotropic polarization of speckle patterns~\cite{schmidt2015isotropically}, to light transport phenomena~\cite{Asymmetry}.

In the recent past, the applications of these materials to enhance chiral sensing and circular dichroism spectroscopy have spawned. First, surface enhanced molecular circular dichroism was predicted on isolated dual particles \cite{PRB}. Soon after, dual metasurfaces, namely, planar arrays of Kerker-like particles designed to control the properties of light at both the far- and the near-field, proved to be more efficient in revealing and enhancing spectroscopic signals of chiral molecules \cite{ACSHo, ACSSolomon, ACSLasa, PRLFeis}. Unfortunately, the spectroscopic features of many industrially relevant molecules appear in the ultraviolet, where most of the HRI materials that constitute the building blocks of the aforementioned metasurfaces present losses~\cite{ACSUVDionne}. In this regime, the study of helicity preserving dual structures is scarce. For example, it has been recently shown that losses inhibit the emergence of the first Kerker condition for homogeneous nonmagnetic spheres~\cite{PRLJorge}. Consequently, the conservation of helicity is precluded for metasurfaces made of lossy spherical resonators, limiting their possible application in chiral spectroscopy. This is the current state-of-the-art of the intertwining between helicity, chirality, and energy conservation.

In this work, we demonstrate that the presence of losses not only precludes the emergence of the first Kerker condition in nonmagnetic spheres. We show that these correlations are more general, affecting also other systems such as core-shells or chiral inorganic spheres. To do so, we first settle an homogenized framework for the simultaneous study of energy and helicity measurements within general electromagnetic scattering theory. We detail a generally applicable procedure that permits the calculation of the normalized helicity expectation value, $\langle \hat{\Lambda} \rangle$, which is the observable that, in practice, permits the identification of helicity preserving scatterers. Then, we apply this method to the case of a chiral inorganic sphere illuminated by a circularly polarized planewave and explicitly obtain a particular expression for $\langle \hat{\Lambda} \rangle$. The formula obtained is a generalization of a previously reported one \cite{PRLJorge}, which allowed for the evaluation of the helicity expectation value in spherical and core-shell particles. Making use of this expression, we finally show that the presence of losses in core-shells and chiral inorganic spheres sets an upper bound to the helicity expectation value.

The paper is organized as follows: in Section 1, we set the general stage for the calculation of the helicity expectation value; Section 2 is devoted to the derivation of the helicity expectation value for a generic chiral sphere; in Section 3 we show that losses impede helicity conservation in core-shells and chiral inorganic spheres; finally, in Section 4 we resume the main conclusions of our work. The interested reader in scattering theory and the general discussion on the helicity expectation value is advised to follow the whole derivation. On the other hand, the reader only interested in the results for helicity preserving structures can go directly to Section 2.

\section{Helicity expectation value for a generic linear electromagnetic scatterer}
\begin{figure}[h]
    \centering
    \includegraphics[scale = 0.45]{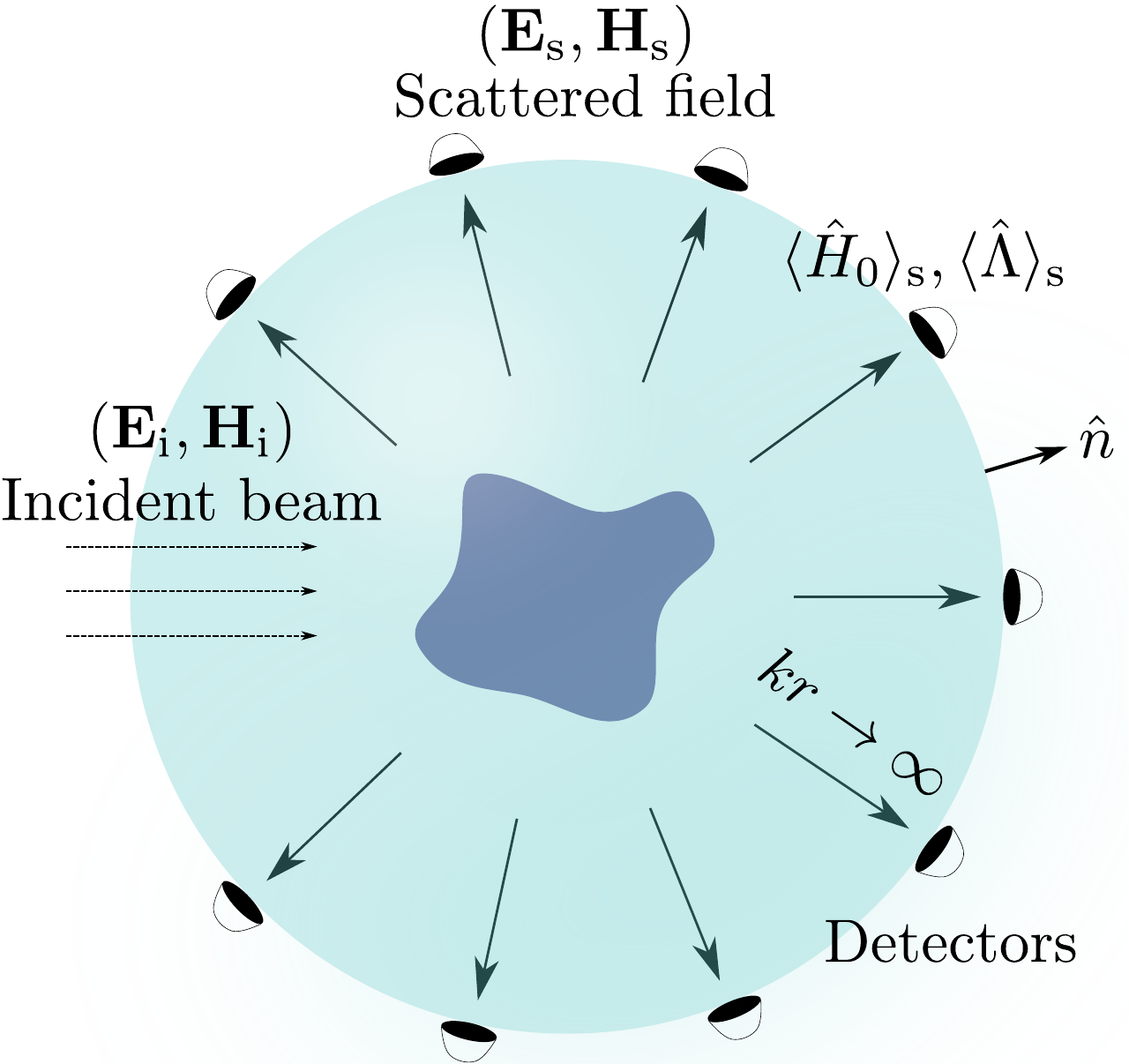}
    \caption{Scheme of a generic electromagnetic scattering measurement setup. An incident electromagnetic field described by $\left( \mathbf{E}_\text{i}, \mathbf{H}_\text{i} \right)$ illuminates a generic sample. In turn, the sample scatters an electromagnetic field $\left( \mathbf{E}_\text{s}, \mathbf{H}_\text{s} \right)$ which is measured by a set of detectors placed in the radiation zone. $\langle \hat{H}_0 \rangle_\text{s}$ and $\langle \hat{\Lambda} \rangle_\text{s}$ are the total scattered energy and helicity. $\hat{n}$ is the unitary normal vector of an imaginary sphere in which the detectors are placed.}
    \label{SetUp}
\end{figure}

First and foremost, let us briefly put forward how measurements are considered within general linear scattering theory. A scheme of a typical scattering measurement setup is given in Fig. \ref{SetUp}. A set of detectors are used to measure the energy and polarization densities. Mathematically, this can be expressed with measurements in the operators $\hat{H}_0$ (hamiltonian) and $\hat{\Lambda}$ (helicity), respectively. Moreover, detectors are usually placed in the far-field region where energy can only exist in the form of electromagnetic waves. In this limit, the total scattered energy per unit time is usually calculated from the flux of the scattered Poynting vector, $\mathbf{S}_\text{s}$, across a spherical surface with normal vector $\hat{n}$ which is centered in the sample. However, one can check that in the radiation zone $\mathbf{S}_\text{s} \cdot \hat{n} = c\langle \hat{H}_0 \rangle$ \cite{Jackson},  where $c$ is the speed of light in vacuum and the usual notation $\langle \hat{O} \rangle$ is employed to express the expected value of an operator $\hat{O}$. Consequently, the flux is proportional to the local energy density in the far-field. In terms of experimentally quantifiable magnitudes, such as Stokes parameters, measuring $\hat{H}_0$ would be associated with evaluating the parameter $s_0$, whereas determining $\hat{\Lambda}$ with the parameter $s_3$ \cite{CrichtonAndMarston, BirulaStokes}. Finally, we consider that measurements are carried in a spherical surface around the sample which covers all the possible scattering angles.

To deal with scattering problems in the most general way it is convenient to introduce the expansions of the scattered electromagnetic field in terms of vector spherical harmonics, as given in several books of reference in the field \cite{Jackson, Zangwill}. The expansion of a monochromatic electric field can be written as (see, for instance, Eq. 9.122 in \cite{Jackson}):
\begin{align}
    &\mathbf{E}_\text{s}(\mathbf{r}) = \sum_{\ell, m} \left[B_ {\ell m}\mathbf{U}_{\ell m}(\mathbf{r}) + i A_ {\ell m} \mathbf{V}_{\ell m}(\mathbf{r}) \right]
    \label{Jack}
\end{align}
where $i = \sqrt{-1}$, $A_{\ell m}$ and $B_{\ell m}$ are the electric and magnetic frequency-dependent scattering coefficients, which most generally depend on the incident illumination; $\mathbf{U}_{\ell m}(\mathbf{r}) = h_\ell(kr)\mathbf{X}_{\ell m}(\theta, \varphi)$ and $\mathbf{V}_{\ell m}(\mathbf{r}) = \frac{1}{k}\nabla\times\mathbf{U}_{\ell m}(\mathbf{r})$ are Hansen's multipoles, with $h_\ell(kr)$ the spherical Hankel function of the first kind and $\mathbf{X}_{\ell m}(\theta, \varphi)$ the vector spherical harmonics \cite{Jackson}; finally, $k$ is the wavenumber of light in vacuum, and a time-harmonic dependence $\exp(-i\omega t)$ is considered through the text, with $\omega$ the frequency of light. Analytical expressions for the electric and magnetic complex scattering coefficients are available only for certain specific problems. In general, $A_{\ell m}$ and $B_{\ell m}$ coefficients can be computed numerically if the current distribution in the interior part of the sample is known~\cite{ExactMultipolar}. The possibility of computing the scattering coefficients from an arbitrary distribution of currents makes the expression in Eq. \eqref{Jack} completely general and applicable to any solvable electromagnetic scattering problem. On the other hand, for our aim it is necessary to impose certain constraints on the multipolar expansion of the incident beam. In particular, we require the incident beam to have a well-defined helicity, $\sigma = \pm 1$. This is an imperative condition if we want to obtain an observable, $\langle \hat{\Lambda} \rangle$, that helps us in the identification of helicity preserving structures. This type of illumination can be generally written as:
\begin{equation}
    \mathbf{E}_i(\mathbf{r}) = \sum_{\ell, m}C^\sigma_{\ell m}\left[ \frac{\Tilde{\mathbf{U}}_{\ell m}(\mathbf{r}) +\sigma \Tilde{\mathbf{V}}_{\ell m}(\mathbf{r})}{\sqrt{2}} \right],
\end{equation}
where $C_{\ell m}^\sigma$ are the expansion coefficients which determine the spatial form of the incident beam. $\Tilde{\mathbf{U}}_{\ell m}(\mathbf{r}) = j_\ell(kr)\mathbf{X}_{\ell m}(\theta, \varphi)$ and $\Tilde{\mathbf{V}}_{\ell m}(\mathbf{r}) = \frac{1}{k}\nabla\times\Tilde{\mathbf{U}}_{\ell m}(\mathbf{r})$, where $j_\ell(kr)$ is the spherical Bessel function of the first kind.

Even if the expression in Eq. \eqref{Jack} fully determines the fields generated from an arbitrary harmonic scatterer, it is more convenient for our aim to switch to the Riemann-Silberstein (RS) representation of the electromagnetic field. The RS vectors are constructed by taking the following superpositions of the complex electric and magnetic fields: $\mathbf{F}^\pm (\mathbf{r}) = \sqrt{\varepsilon_0/2}[\mathbf{E} (\mathbf{r}) \pm iZ_0\mathbf{H} (\mathbf{r})]$, where $\varepsilon_0$ is the permittivity of vacuum and $Z_0$ is the impedance of vacuum. The column vector $\mathcal{F} = \{\mathbf{F}^+ (\mathbf{r}),  \mathbf{F}^- (\mathbf{r})\}$ has been elsewhere regarded as the object from which the expectation values associated to fundamental electromagnetic observables can be obtained \cite{Birula1}. Indeed, the time-averaged local density of electromagnetic energy can be derived as $\langle \hat{H}_0 (\mathbf{r}) \rangle = \frac{1}{2\omega}\mathcal{F}^\dag \hat{H}_0 \mathcal{F} = \frac{1}{2}\mathcal{F}^\dag \mathcal{F}$, where the "$\dag$" symbol means transpose conjugate and $\hat{H}_0 = i\partial_t$. The time-averaged local density of electromagnetic helicity can also be computed in this notation as $\langle \hat{\Lambda} (\mathbf{r}) \rangle = \frac{1}{2\omega} \mathcal{F}^\dag \hat{\Lambda} \mathcal{F}$, bearing in mind that the RS vectors are eigenstates of the helicity operator, $\hat{\Lambda}\mathbf{F}^\pm(\mathbf{r}) = \pm \mathbf{F}^\pm (\mathbf{r})$, with $\hat{\Lambda} = \frac{1}{k}\nabla\times$. Moreover, let us define the ratio between the scattering electric/magnetic multipolar coefficients and the incident electric/magnetic coefficients as:  $\alpha_{\ell m} \equiv i\sigma\sqrt{2}A_{\ell m}/C_{\ell m}^\sigma$ and $\beta_{\ell m} \equiv \sqrt{2}B_{\ell m}/C_{\ell m}^\sigma$. As a result, in the RS representation, the field scattered by an arbitrary sample which has been illuminated by an incident beam with well-defined helicity $\sigma$ can be compactly written as:
\begin{equation}
    \mathbf{F}_\text{s}^{\sigma'} (\mathbf{r}) = \sqrt{\varepsilon_0}\sum_{\ell, m} C_{\ell m}^\sigma \left[\frac{\beta_ {\ell m} + \sigma\sigma' \alpha_ {\ell m}}{\sqrt{2}} \right] \left[ \frac{\mathbf{U}_{\ell m}(\mathbf{r}) +\sigma' \mathbf{V}_{\ell m}(\mathbf{r})}{\sqrt{2}} \right],
    \label{RSgeneral}
\end{equation}
where $\sigma'=\pm 1$ defines the different helicity components of the scattered field.

Determining the optical scattering of an object implies measuring the total radiation over the full solid angle of $4\pi$ at large distances from the sample under study. Experimentally, this is usually achieved with the aid of an integrating sphere. Mathematically, in order to calculate the total scattered energy one needs to integrate the observable $\langle \hat{H}_0(\mathbf{r}) \rangle$ over the full solid angle. Similarly, the computation of the total scattered helicity requires to integrate the observable $\langle \hat{\Lambda}(\mathbf{r}) \rangle$ over all possible scattering directions. In addition, one can split the scattered fields into its helicity conserved ($\sigma'=+\sigma$) and helicity flipped ($\sigma' = -\sigma$) components. Making use of this notation, one can evaluate the energy and helicity densities for the whole solid angle:
\begin{align}
    \label{SolAngleInt1}
    \langle \hat{H}_0 (r) \rangle &= \frac{1}{2}\int \left( |\mathbf{F}_\text{s}^{+\sigma}(\mathbf{r})|^2 + |\mathbf{F}_\text{s}^{-\sigma}(\mathbf{r})|^2 \right) d\Omega\\
    \langle \hat{\Lambda} (r) \rangle &= \frac{1}{2\omega}\int \left( |\mathbf{F}_\text{s}^{+\sigma}(\mathbf{r})|^2 - |\mathbf{F}_\text{s}^{-\sigma}(\mathbf{r})|^2 \right) d\Omega,
    \label{SolAngleInt2}
\end{align}
where $d\Omega = \sin\theta d\theta d\varphi$, the integration limits being $\theta\in[0, \pi]$ and $\varphi \in[0, 2\pi]$. Employing the orthogonality properties of the vector spherical harmonics $\mathbf{U}_{\ell m}$ and $\mathbf{V}_{\ell m}$ \cite{Barrera, Carrascal}, one can explicitly calculate the energy and helicity densities as a function of the radial coordinate, $r$. However, as previously stated, in scattering theory measurements are typically carried out in the radiation zone. This implies that the total scattered energy and helicity observables are obtained by integrating only the far-field terms. Analytically, this means that the total scattered energy and helicity are obtained as:
\begin{align}
    \label{H0scat}
    \langle \hat{H}_0 \rangle_\text{s} &= \lim_{kr \rightarrow \infty} (kr)^2 \langle \hat{H}_0 (r) \rangle = \frac{\varepsilon_0}{2} \sum_{\ell, m} |C_{\ell m}^\sigma|^2\left( |\alpha_ {\ell m}|^2 + |\beta_ {\ell m}|^2 \right)\\
    \label{Lscat}
    \langle \hat{\Lambda} \rangle_\text{s} &= \lim_{kr \rightarrow \infty} (kr)^2 \langle \hat{\Lambda} (r) \rangle = \frac{\varepsilon_0}{\omega} \sum_{\ell, m} |C_{\ell m}^\sigma|^2\Re\left(\alpha_ {\ell m}^* \beta_{\ell m} \right),
\end{align}
where $\Re$ represents the real part of a complex number and in the derivation we have used the fact that $\lim_{kr \rightarrow \infty} (kr)^2|h_\ell|^2 = \lim_{kr \rightarrow \infty} |\frac{\partial}{\partial r} (r h_\ell)|^2 = 1$. From Eq. \eqref{H0scat} and Eq. \eqref{Lscat} we can derive a completely general expression for the normalized (and unit-less) helicity expectation value:
\begin{equation}
    \langle \hat{\Lambda} \rangle \equiv \frac{\omega\langle \hat{\Lambda} \rangle_\text{s}}{\langle \hat{H}_0 \rangle_\text{s}} = 2 \frac{\sum_{\ell, m} |C_{\ell m}^\sigma|^2 \Re(\alpha_{\ell m}^*\beta_{\ell m})}{\sum_{\ell, m}|C_{\ell m}^\sigma|^2\left( |\alpha_{\ell m}|^2 + |\beta_{\ell m}|^2 \right)}.
    \label{Lambda}
\end{equation}

\begin{figure}[t]
    \centering
    \includegraphics[scale = 0.35]{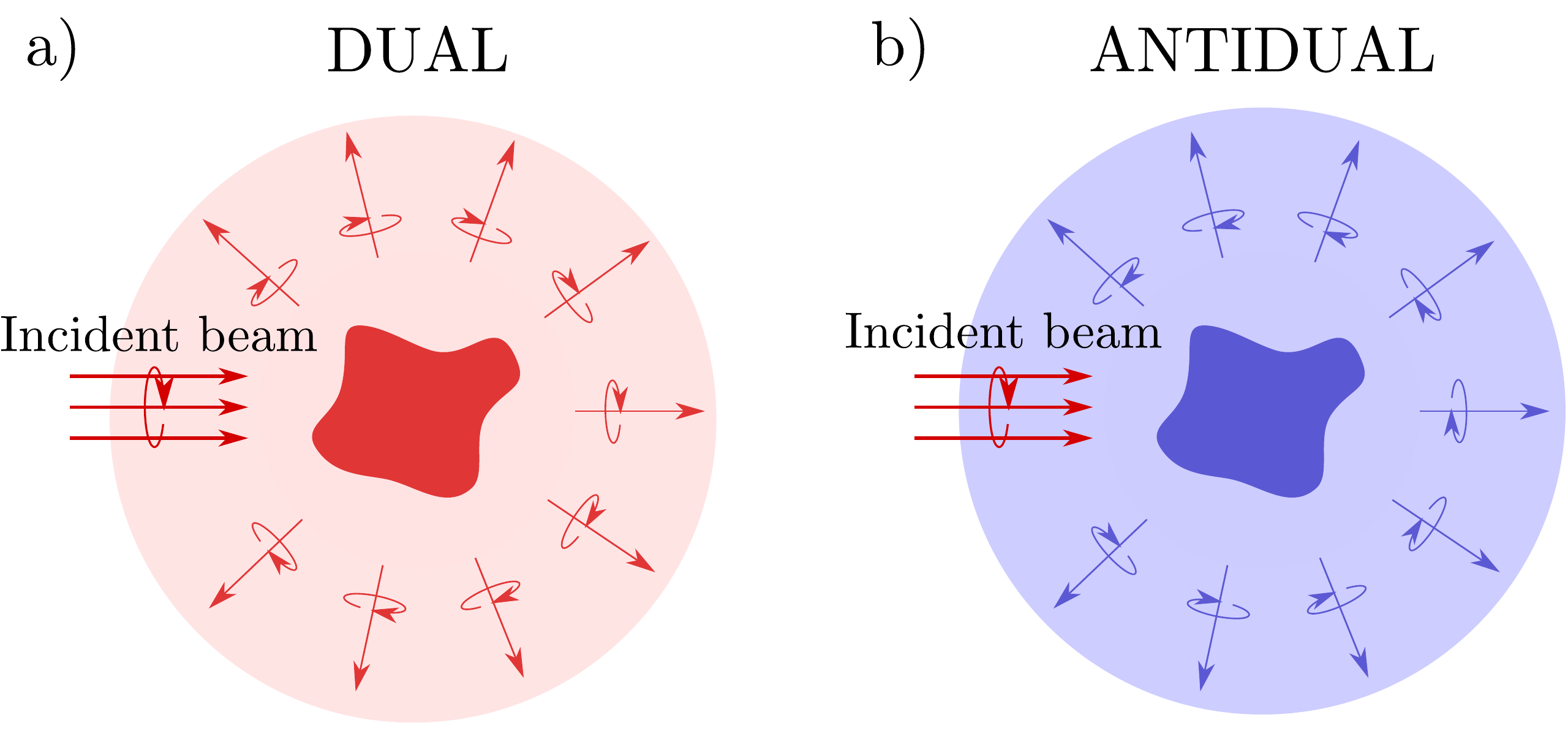}
    \caption{a) A sketch of a generic dual sample, i.e. a scatterer which preserves the helicity of an incident electromagnetic field; b) a sketch of a generic antidual sample, i.e. a scatterer which flips the helicity of an incident electromagnetic field. Red and blue arrows represent circularly polarized planewaves with opposite helicities.}
    \label{DualAntidual}
\end{figure}

This is a physical magnitude which can also be determined in terms of the experimentally quantifiable Stokes parameters: $\langle \hat{\Lambda} \rangle = \sigma \int s_3 d\Omega/ \int s_0 d\Omega$. It is a bounded observable, $\langle \hat{\Lambda} \rangle \in [-1, 1]$, and its extreme values correspond to a dual, $\langle \hat{\Lambda}\rangle = +1$, or an antidual, $\langle \hat{\Lambda}\rangle = -1$, scatterer. Dual scatterers are those which preserve the helicity of the incident beam (Fig. \ref{DualAntidual}a), whereas antidual samples are those which flip it (Fig. \ref{DualAntidual}b). Given an incident electromagnetic beam with well-defined helicity $\sigma$, a dual scatterer is defined as one which only emits in the same helicity components ($\mathbf{F}_\text{s}^{-\sigma} = 0$). On the contrary, an antidual sample only scatters in the opposite helicity components ($\mathbf{F}_\text{s}^{+\sigma} = 0$). As it can be checked from the expression in Eq. \eqref{RSgeneral}, dual scatterers are generally found whenever $\alpha_{\ell m} = \beta_{\ell m}$, whereas antidual scatterers are obtained whenever $\alpha_{\ell m} = -\beta_{\ell m}$. Importantly, we would like to highlight once again that the expression given in Eq. \eqref{Lambda} helps in the identification of dual or antidual scatterers if and only if the incident beam has a well-defined helicity $\sigma$. In the case of the incident beam been a superposition of helicity states (for instance, a linearly polarized planewave) the scattered field is generally described by both RS vector components and, consequently, $\langle \hat{\Lambda} \rangle$ loses its usefulness in the identification of helicity preserving scatterers.

Finally, to finish with the discussion of the general expression for $\langle \hat{\Lambda} \rangle$ given in Eq. \eqref{Lambda}, we would like to note that the multipolar basis chosen in Eq. \eqref{Jack} is not the only one that can be employed for the derivation. Any other complete basis set may be adopted if one knows the solid angle integral orthogonality expressions and the limits in the far-field. Moreover, we would also like to note that the normalized helicity expectation value can also be computed numerically from the first equality in Eq. \eqref{Lambda}: given a point by point solution of the scattered electromagnetic field, one can always compute the integrals in Eq. \eqref{SolAngleInt1} and Eq. \eqref{SolAngleInt2} in the radiation zone as discrete sums. This makes our approach to calculate $\langle \hat{\Lambda} \rangle$ completely general and applicable to any electromagnetic scattering problem for which the solution of the scattered fields is known.

\section{Helicity expectation value for a chiral inorganic sphere}
As a practical implementation of all the previously discussed ideas, we have calculated the normalized helicity expectation value for an optically active (or chiral) inorganic spherical sample. Note that such chiral spheres can capture the optical response of artificial chiral systems \cite{PRX_ArtificialChiral, NanoLetters_ArtificialChiral} in an effective medium --- orientation averaged --- approximation. The light scattering by a chiral sphere under planewave illumination has a closed analytical solution (see Section 8.3 in Ref. \cite{Bohren}). The multipolar expansion of such scattered fields is actually a generalization of  the well-known Mie theory for spherical particles and can be expressed as:
\begin{align}
    \nonumber
    \mathbf{E}_\text{s} (\mathbf{r}) &=  \sum_{\ell=1}^\infty E_\ell \left\{ i a_\ell\mathbf{N}_{e 1 \ell}^{(3)} - b_\ell\mathbf{M}_{o 1 \ell}^{(3)} + c_   {\ell}\left(\mathbf{M}_{e 1 \ell}^{(3)} + i \mathbf{N}_{o 1 \ell}^{(3)}\right) \right\}\\
    iZ_0\mathbf{H}_\text{s}(\mathbf{r}) &= \sum_{\ell=1}^\infty E_\ell \left\{ i a_\ell\mathbf{M}_{e 1 \ell}^{(3)} - b_\ell\mathbf{N}_{o 1 \ell}^{(3)} +c_   {\ell}\left(\mathbf{N}_{e 1 \ell}^{(3)} + i \mathbf{M}_{o 1 \ell}^{(3)}\right) \right\}
    \label{ScaChiral}
\end{align}
where $E_\ell = E_0i^\ell (2\ell + 1)/\ell(\ell + 1)$ and $E_0$ is the amplitude of the incident electric electric field. $\{\mathbf{M}_{e |m| \ell}, \mathbf{M}_{o |m| \ell}, \mathbf{N}_{e |m| \ell}, \mathbf{N}_{o |m| \ell}\}$ are the vector spherical harmonics used by Stratton (see Section 7.11 in Ref. \cite{Stratton}) and Bohren, which can also be related to the set employed by Jackson \cite{NoraMultipoles}. Finally, $a_\ell$, $b_ \ell$ and $c_\ell$ are the Mie coefficients which can be computed either for an optically active sphere (see page 188 in \cite{Bohren}), a conventional sphere (see Eq. 4.56 and Eq. 4.57 in Ref. \cite{Bohren}) or a core-shell spherical system (see Eq. 8.2 in Ref. \cite{Bohren}) requiring, in these last two cases, that $c_\ell = 0$.

\begin{figure*}[t]
    \centering
    \includegraphics[width=\textwidth]{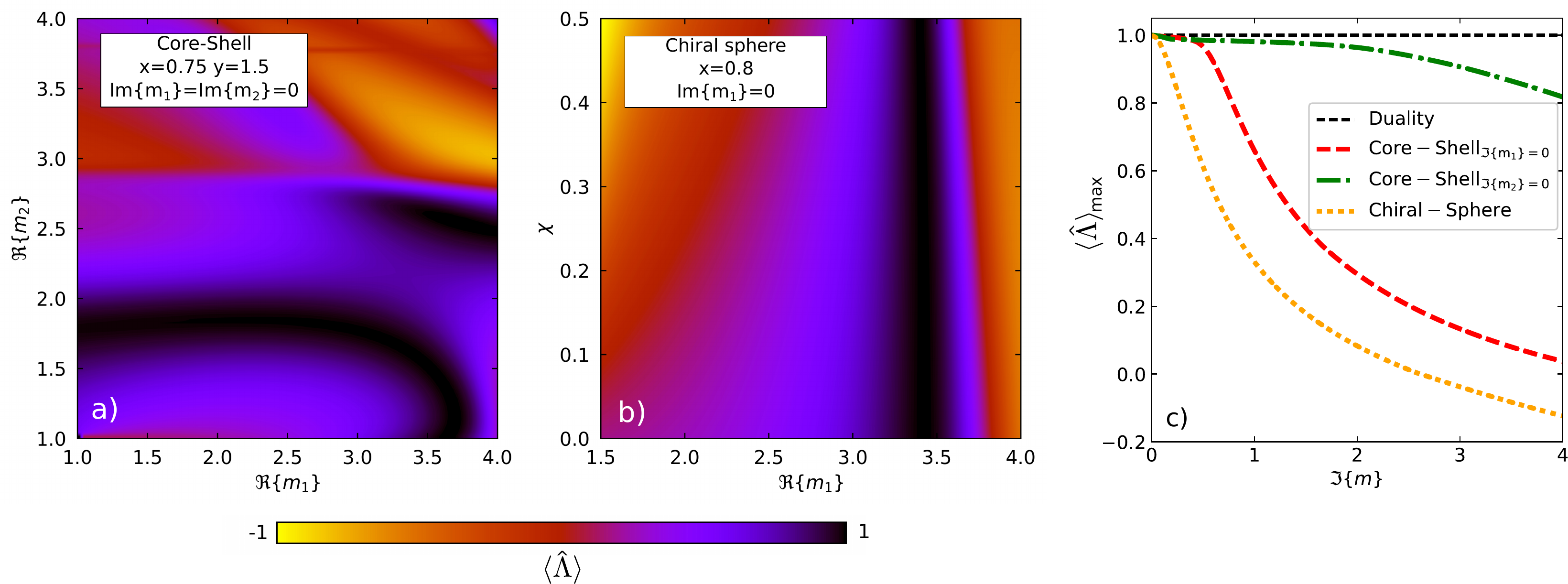}
    \caption{Helicity expectation value, $\langle \hat{\Lambda} \rangle$, for a core-shell and an optically active particle. a) Helicity expectation value for a lossless multipolar core-shell with inner size parameter $x = 0.75$ and outer size parameter $y = 1.5$. b) Helicity expectation value for a lossless spherical chiral particle with a size parameter $x = 0.8$ illuminated with a left polarized planewave. c) Maximum of the average helicity value, $\langle \hat{\Lambda} \rangle_\text{max}$, when modifying $\Im(m_2)$ for the core-shell system (red line), $\Im(m_1)$ for the core-shell system (green line) and $\Im(m_1)$ for the chiral particle (orange line).}
    \label{fig_3}
\end{figure*}

It is of utmost importance to note that the incident field employed to compute the scattered fields in Eq. \eqref{ScaChiral} is a linearly polarized planewave propagating along the $OZ$ direction (see Eq. 4.21 and Eq. 4.37 in Ref. \cite{Bohren}). Thus, to obtain a meaningful expression of the normalized helicity expectation value, we should consider only one circular polarization component for the incident field and, consequently, take into account only the scattered fields associated to such a component. We have done this by switching from Bohren's vector spherical harmonics notation in Eq. \eqref{ScaChiral} to Jackson's notation and identifying the different total angular momentum contributions, $m = \pm 1$, to the expansion of the incident planewave. As a planewave contains no orbital angular momentum, $m$ is fully associated to the z-component of the spin angular momentum and, therefore, to the two possible circular polarizations (or helicities). Moreover, as $m$ is the eigenvalue associated to the z-component of the total angular momentum operator, it is a conserved quantity in scattering problems with cylindrically symmetric samples along the $OZ$ axis. Thus, to identify the fields scattered due to a single helicity component of the incident field, we must separately consider the terms in the expansion given by Eq. \eqref{ScaChiral} with a total angular momentum $m = 1$ and $m = -1$.

By doing so and following, step by step, the previously discussed general procedure, one obtains a closed analytical multipolar expression of the normalized helicity expectation value for a chiral sphere:
\begin{equation}
    \langle \hat{\Lambda} \rangle = 2\frac{\sum_{\ell = 1}^\infty (2\ell + 1)\left\{ \Re(a_\ell^*b_\ell) + |c_\ell|^2\ + \sigma\Im \left[(a_\ell + b_\ell)^*c_\ell\right] \right\}}{\sum_{\ell = 1}^\infty (2\ell + 1)\left\{ |a_\ell|^2 + |b_\ell|^2 + 2|c_\ell|^2 + 2 \sigma \Im \left[(a_\ell + b_\ell)^*c_\ell\right] \right\}},
    \label{LambdaChiral}
\end{equation}
where, $\Im$ represents the imaginary part of a complex number. It is worth noting that for the cases of a sphere and a core-shell, i.e. whenever $c_\ell = 0$, one recovers the well-known expression of the normalized helicity expectation value for such systems \cite{PRLJorge, Unveiling}. However, considering the $c_\ell$ coefficients makes the expression in Eq. \eqref{LambdaChiral} also applicable to optically active spheres. It can be checked that, in opposition to the case of a sphere and core-shell, this more general expression for $\langle \hat{\Lambda} \rangle$ critically depends on the helicity of the incident field, $\sigma$. This is so because, by definition, chiral scatterers are those whose response depends on the polarization of the incident beam, i.e. their refractive index is a function of the input helicity state. Expressions similar to Eq. \eqref{LambdaChiral} can be calculated for other systems for which an analytical solution is known, for instance, an optically active spherical shell \cite{OpticallyActiveShell}.

In what follows, we employ the expression for the helicity expectation value in Eq. \eqref{LambdaChiral} to show that the correlations between energy and helicity reported in Ref. \cite{PRLJorge} are also present in core-shells and optically active spheres. Therein it was analytically demonstrated that helicity preservation in nonmagnetic spheres, the so-called first Kerker condition, can only be achieved for scatterers built up from lossless materials. Moreover, it is also known that the helicity conservation condition can only be achieved in pure multipolar regimes, i.e. in spectral areas where only one multipolar order $\ell$ dominates \cite{Unveiling}. In practice, dual particles are usually constructed within the dipolar regime ($\ell = 1$) or, in other words, in the limit where the particle is small compared to the illuminating wavelength. Consequently, the reported results imply that dipolar helicity preserving spheres can only be constructed with materials and at wavelengths in which the refractive index, $m = \sqrt{\varepsilon}$, is a real number. This was shown by using analytical properties of the Bessel functions that define the electric and magnetic Mie coefficients, which indicates that the correlations might be also present in other nonmagnetic scatterers. With the aid of Eq. \eqref{LambdaChiral}, we show that these correlations are indeed more general.

\section{Helicity conservation in lossy core-shells and chiral inorganic spheres}
To show that the correlations mentioned above between energy and helicity are not limited to the dipolar regime, we have first addressed the case of a multipolar core-shell, i.e. a core-shell that cannot be described only by the first Mie coefficients $a_1$ and $b_1$. For core-shell systems under planewave illumination, the expression for $\langle \hat{\Lambda} \rangle$ is exactly the same as the one found for spheres, i.e. imposing $c_\ell = 0$ in Eq. \eqref{LambdaChiral} and substituting the appropriate analytical expressions for the scattering coefficients $a_\ell$ and $b_\ell$. In Fig. \ref{fig_3}a we show the computed normalized helicity expectation value for a wide range of real refractive indices associated both to the core, $m_1$, and to the shell, $m_2$. The size parameters of the core, $x = ka$, and the shell, $y = mx$, have been chosen such that the response of the system is determined both by the dipolar ($a_1$, $b_1$) and quadrupolar ($a_2$, $b_2$) Mie coefficients. As it can be checked from Fig. \ref{fig_3}a, when setting the imaginary parts of the refractive indices to zero, there are several combinations of $\Re(m_1)$ and $\Re(m_2)$ which enable the construction of helicity preserving scatterers, i.e. the points at which $\langle \hat{\Lambda} \rangle = +1$ (black areas).

On the other hand, we have also computed the normalized helicity expectation value for a system in which the constituent relations consider cross electric-magnetic terms, i.e. media in which electric fields generate magnetization currents and magnetic fields generate polarization currents. In Fig. \ref{fig_3}b we show the helicity expectation value for a system characterized by a chiral constant $\chi$, which determines the refractive indices of a chiral medium both for a left ($\sigma = +1$) or right ($\sigma = -1$) polarized incident planewave. More explicitly, the refractive indices are computed as $n_\sigma = m_1 - \sigma\chi$. As it can be noted from the figure, for a dipolar particle ($x = 0.8$), helicity preserving samples can be constructed for several values of $\chi$ and with an approximately constant value of $\Re(m_1)$. For an optically active dipolar sphere, the chiral nature of the media induces a symmetric change in the electric, $a_1$, and magnetic, $b_1$, dipolar scattering coefficients. Given an incident circularly polarized planewave, the electric scattering coefficient is effectively changed into $a_1^{ch} = a_1 - i\sigma c_1$ and, similarly, the magnetic scattering coefficient into $b_1^{ch} = b_1 - i\sigma c_1$. Thus, starting at the first Kerker condition for a nonchiral spherical particle ($\chi = 0$), the equality of the electric and magnetic dipolar scattering coefficients is seen to hold well also for points at which $\chi \neq 0$.

In Fig. \ref{fig_3}c, we analyze the behaviour of core-shells and chiral particles when losses are considered. To do so, we have progressively modified the imaginary part of the refractive indices, $\Im(m)$, recalculating the maps equivalent to the ones shown in Fig. \ref{fig_3}a and Fig. \ref{fig_3}b. For each value of the imaginary part of the refractive index, we take the maximum value of the whole colormap, $\langle \hat{\Lambda} \rangle_\text{max}$, and then we plot it against $\Im(m)$ in Fig. \ref{fig_3}c. As it can be checked, all the maximum values of the computed helicity expectation values are sensitive to the presence of absorption in the system. For the case of the core-shell, introducing losses in the outer volume leads to a much faster decay of $\langle \hat{\Lambda} \rangle_\text{max}$ (red line). However, the system appears to be more robust to the presence of losses in the inner volume, even if still $\langle \hat{\Lambda} \rangle_\text{max} < 1$ for any value of $\Im(m) > 0$ (green line). This result is in agreement with previous results reported for core-shells in Ref.\cite{WeiLiu}. For core-shells with small core radii, the presence of losses in the inner volume does not significantly increase the absorption cross-section of the system as a whole. Thus, as it was shown therein, unidirectionality of the scattered field, which implies helicity conservation for cylindrically symmetric structures, can still be approximately reached in this case.

\section{Conclusions}
In conclusion, we have set the general stage for the study of helicity conservation from the observational point of view in arbitrary electromagnetic scattering problems. We have given an expression for the normalized helicity expectation value, i.e. Eq. \eqref{Lambda}, which permits the identification of helicity preserving structures in the most possible general way. Then, we have applied the general procedure to a case of practical interest, i.e the fields scattered by a chiral inorganic sphere, obtaining a closed analytical expression for $\langle \hat{\Lambda} \rangle$ in Eq. \eqref{LambdaChiral}. Finally, we have employed this expression to show that correlations between helicity and energy preservation are also present in chiral inorganic spheres and core-shells. On the one hand, we have shown that losses in chiral spheres and in the outer volume of core-shells completely preclude the first Kerker condition. On the other hand, losses in the interior part of core-shell systems seem to be less critical for the preservation of helicity, but still determinant. These results indicate that lossy nonmagnetic materials are not ideal candidates to construct helicity preserving metasurfaces and, thus, to build suitable systems to enhance CD spectroscopy of surrounding chiral molecules.

\section*{Author Contributions}

J.L.A., J.O.T., A.G.E., and G.M.T. initiated the project. J.L.A. and J.O.T. developed the theory and performed the simulations. All the authors discussed the results and wrote the manuscript. A.G.E. and G.M.T. coordinated the project.

\section*{Conflicts of interest}

There are no conflicts to declare.

\section*{Acknowledgements}

J.L.A., J.O.T. and A.G.E. acknowledge support from the Spanish Ministerio de Ciencia e Innovación (PID2019-109905GA-C2) and from Eusko Jaurlaritza (IT1164-19, KK-2019/00101 and KK-2021/00082). A.G.E. acknowledges funding from Programa Red Guipuzcoana de Ciencia, Tecnología e Innovación 2021 (Grant Nr. 2021-CIEN-000070-01. Gipuzkoa Next). A.G.E. and G.M.T. acknowledge funding from the Basque Government's IKUR initiative on Quantum technologies (Department of Education). The authors would like to acknowledge insightful discussions with Dr. Nuno de Sousa. J.L.A. would also like to thank Carlos Maciel Escudero for valuable discussions on the Riemann-Silberstein vectors and Álvaro Nodar for his indications regarding the choice of vector spherical harmonics.



\balance


\bibliography{mybib} 
\bibliographystyle{rsc} 

\end{document}